# A new theory of fluid-solid coupling in a porous medium for application to the ultrasonic evaluation of tissue remodeling using bioelastomers

**Chuanyang Jiang[a], Yanying Zhu[b], Kaixuan Guo[b], Qing Li[b], Zhengwei You[c], Jiao Yu[b,*]**

[a]College of Mechanical Engineering, Liaoning Shihua University, Fushun, Liaoning Province, 113001, P. R. China

[b]College of Science, Liaoning Shihua University, Fushun, Liaoning Province, 113001, P. R. China

[c]State Key Laboratory for Modification of Chemical Fibers and Polymer Materials, Shanghai Belt and Road Joint Laboratory of Advanced Fiber and Low-dimension Materials (Donghua University), College of Materials Science and Engineering, Donghua University, Shanghai, 201620, P. R. China

---

[*] Corresponding author. Address: College of Science, Liaoning Shihua University, Fushun, Liaoning Province, 113001, P. R. China. Email: yujiaojoy@hotmail.com. Tel.: 86-24-56865706. Fax: 86-24-56860766.

# ABSTRACT

Bioelastomers have demonstrated tremendous value and potential in the field of tissue repair due to increasing health demands. Improved non-invasive methods are required for monitoring tissue development assisted by bioelastomers. In this paper, we present a novel theory of fluid-solid coupling in a porous medium for application to the ultrasonic evaluation of tissue remodeling using bioelastomers. The common assumption of equal solid and liquid displacements used in the conventional description of a fluid-saturated porous solid cannot be applied to soft media, such as bioelastomers. We revise the geoacoustic theory of Biot to allow for relative motion between a fluid and a solid in an aggregate and derive an expression for a characteristic fluid-solid coupling parameter. Unlike the conventional method, the propagation speed of shear waves observed by ultrasound shear wave elastography is considered a known quantity in the novel theory, and the calculated value of the coupling parameter is used to evaluate the status of tissue repair. The model is validated by analyzing selected cases. The conditions under which the model can be applied are identified. However, further development of the theory is required to extract dynamic parameters that can be used to monitor the entire tissue remodeling process. In this paper, a theoretical approach is developed that can be used to analyze the mechanics of tissue repair. The theory has potential applications in the field of acellular in situ tissue engineering for non-invasive monitoring of the complex mechanical remodeling process of tissue regeneration and bioelastomer degradation.

# 1. INTRODUCTION

The rapid development of acellular in situ tissue engineering for organ and tissue repair has created an increasing demand for improved capabilities of non-invasive monitoring of the tissue regeneration and repair process. Porous bioelastomers at the implant site perform an important mechanical function in the tissue remodeling process in that they support tissue in-growth before being degraded, until the implant site is eventually replaced with native tissue.

Based on our previous experiments and studies (Yu et al., 2013; Dutta et al., 2012; Yu et al., 2018), we build a theoretical model to evaluate the mechanical properties of porous bioelastomers during tissue remodeling and provide guidance for using ultrasound-based technologies to non-invasively monitor the process.

In 1956, Biot (Biot, 1956a; Biot, 1956b) put forward a fundamental theory of the propagation of elastic waves in a fluid-saturated porous solid for geoacoustic studies. A remarkable assumption in Biot's paper is that there is no relative motion between a fluid and a solid in an aggregate. This assumption was used to derive the shear wave velocity. Our current research on implanted bioelastomers shows that a fluid-solid aggregate is composed of soft porous biomaterials and saturated fluids, such that the displacement relations do not satisfy Biot's assumption. Here, based on our previous experiments and an appropriate displacement assumption, we develop a novel theoretical model for the tissue remodeling process.

The model may provide useful mechanical correlations for soft-tissue engineering applications. The shear wave speed is an important mechanical parameter that has become acoustically measurable (Parker et al., 2011; Sigrist et al., 2017; Toyoshima et al., 2020). The developed model may offer novel insights into the application of ultrasound elasticity imaging in the field of tissue engineering. The developed model can be used to meet the challenges encountered in the non-invasive monitoring of the tissue remodeling process in porous bioelastomers.

## 2. METHODS

A fundamental distinction from Biot's model (Biot, 1956a; Biot, 1956b) is that the subjects of our study are soft porous bioelastomers that function as tissue constructs by interacting with fluids inside the body. We replace the assumption of no relative motion between a solid and a fluid in Biot's classical theory with the following relationship:

$$U_x = (1 + \text{n})u_x, \qquad (1)$$

where the solid and the fluid displacements are denoted by $\vec{u}$ and $\vec{U}$, respectively. The softness of porous bioelastomers is taken into account in the equation above. The fluid dynamics are demonstrated by only considering variations in the x-direction for clarity. Note that although the maximum change in the fluid profile may occur in other directions, the simplification provided by selecting the x-direction enables us to connect our theory with Biot's theory.

The kinetic energy T of the fluid-solid aggregate per unit volume is expressed as

$$T = \frac{1}{2}\rho_{11}\left[\left(\frac{\partial u_x}{\partial t}\right)^2 + \left(\frac{\partial u_y}{\partial t}\right)^2 + \left(\frac{\partial u_z}{\partial t}\right)^2\right]$$
$$+\rho_{12}\left[\frac{\partial U_x}{\partial t}\frac{\partial u_x}{\partial t} + \frac{\partial U_y}{\partial t}\frac{\partial u_y}{\partial t} + \frac{\partial U_z}{\partial t}\frac{\partial u_z}{\partial t}\right] \quad (2)$$
$$+\frac{1}{2}\rho_{22}\left[\left(\frac{\partial U_x}{\partial t}\right)^2 + \left(\frac{\partial U_y}{\partial t}\right)^2 + \left(\frac{\partial U_z}{\partial t}\right)^2\right],$$

where the coefficients $\rho_{11}$, $\rho_{12}$ and $\rho_{22}$ are mass coefficients used to incorporate the non-uniformity of the relative motion. For a statistically homogeneous and isotropic porous medium, we may focus on one specific direction without loss of generality. Considering the x-direction, the kinetic energy can be expressed as

$$T = \frac{1}{2}\rho_{11}\left(\frac{\partial u_x}{\partial t}\right)^2 + \rho_{12}(1+n)\left(\frac{\partial u_x}{\partial t}\right)^2 + \frac{1}{2}\rho_{22}(1+n)^2\left(\frac{\partial u_x}{\partial t}\right)^2$$
$$= \frac{1}{2}[\rho_{11} + 2(1+n)\rho_{12} + (1+n)^2\rho_{22}]\left(\frac{\partial u_x}{\partial t}\right)^2. \quad (3)$$

Eq. (3) can be used to obtain the total mass density of the fluid-solid aggregate $\rho$ as

$$\rho = \rho_{11} + 2(1+n)\rho_{12} + (1+n)^2\rho_{22}. \quad (4)$$

Alternatively, $\rho$ can be expressed in terms of the porosity $\emptyset$, the mass density of the solid $\rho_s$, and the mass density of the fluid $\rho_f$. The mass of the solid per unit volume of the aggregate can be expressed as

$$\rho_1 = \rho_s(1-\emptyset), \quad (5)$$

and the mass of the fluid per unit volume of the aggregate can be expressed as

$$\rho_2 = \rho_f\emptyset, \quad (6)$$

which can be combined to yield

$$\rho = \rho_1 + \rho_2 = \rho_s + \emptyset(\rho_f - \rho_s). \quad (7)$$

Taking into account the pressure difference in the fluid, we have

$$-\emptyset\left(\frac{\partial P}{\partial x}\right) = \emptyset\rho_f\left(\frac{\partial^2 U_x}{\partial t^2}\right), \quad (8)$$

where the left-hand side of the equation above is the total force $Q_x$ acting on the fluid per unit volume in the x-direction. We obtain

$$Q_x = \rho_2(1+n)\frac{\partial^2 u_x}{\partial t^2}. \tag{9}$$

Considering the general momentum in Lagrange's equations yields

$$q_x = \frac{\partial}{\partial t}\left(\frac{\partial T}{\partial \dot{u}_x}\right) = \frac{\partial^2}{\partial t^2}(\rho_{11}u_x + \rho_{12}U_x), \tag{10}$$

$$Q_x = \frac{\partial}{\partial t}\left(\frac{\partial T}{\partial \dot{U}_x}\right) = \frac{\partial^2}{\partial t^2}(\rho_{12}u_x + \rho_{22}U_x), \tag{11}$$

where $q_x$ is the total force acting on the solid per unit volume in the x-direction. We can use Eq. (1) to express $Q_x$ as

$$Q_x = [\rho_{12} + (1+n)\rho_{22}]\frac{\partial^2 u_x}{\partial t^2}. \tag{12}$$

Comparing equations (9) and (12) yields

$$\rho_2 = \frac{1}{n+1}\rho_{12} + \rho_{22}. \tag{13}$$

Equations (4) and (7) can then be used to obtain

$$\rho_1 = [\rho_{11} + 2(1+n)\rho_{12} + (1+n)^2\rho_{22}] - \left(\frac{1}{n+1}\rho_{12} + \rho_{22}\right)$$

$$= \rho_{11} + \frac{2n^2+4n+1}{n+1}\rho_{12} + (n^2+2n)\rho_{22}. \tag{14}$$

Furthermore, the total effective mass density of the solid moving in the fluid is expressed as

$$\rho_{11} = \rho_1 + \rho_a, \tag{15}$$

where the total effective mass is equal to the actual mass of the solid plus an additional mass due to the fluid. Similarly, the total effective mass density of the liquid is

$$\rho_{22} = \rho_2 + \rho_a. \tag{16}$$

Substituting Eq. (16) into Eq. (13) yields

$$\rho_a = -\frac{1}{n+1}\rho_{12}. \tag{17}$$

Hence, equations (15)–(17) provide the dynamic relations between the coefficients.

The following partial differential equations were formulated by Biot (Biot, 1956a):

$$N\vec{\nabla}^2\vec{u} + \vec{\nabla}[(A + N)e + Q\varepsilon] = \frac{\partial^2}{\partial t^2}(\rho_{11}\vec{u} + \rho_{12}\vec{U}), \qquad (18)$$

$$\vec{\nabla}(Qe + R\varepsilon) = \frac{\partial^2}{\partial t^2}(\rho_{12}\vec{u} + \rho_{22}\vec{U}), \qquad (19)$$

where e and ε are the normal strain in the solid and the strain in the fluid respectively. A, N, Q and R are the coefficients in the stress-strain relations, where N represents the shear modulus of the biomaterial. Considering that a porous biomaterial is statistically isotropic and homogeneous, the shear waves in this fluid-solid aggregate are independent of compressional waves.

Applying the curl operation to equations (18) and (19) yields

$$\frac{\partial^2}{\partial t^2}(\rho_{11}\vec{\omega} + \rho_{12}\vec{\Omega}) = N\nabla^2\vec{\omega}, \qquad (20)$$

$$\frac{\partial^2}{\partial t^2}(\rho_{12}\vec{\omega} + \rho_{22}\vec{\Omega}) = 0, \qquad (21)$$

where $\vec{\omega} = \text{curl } \vec{u}$, and $\vec{\Omega} = \text{curl } \vec{U}$.

Combining equations (20) and (21) yields

$$\frac{\partial^2\vec{\omega}}{\partial t^2} - \left(\frac{N}{\rho_{11} - \frac{\rho_{12}^2}{\rho_{22}}}\right)\nabla^2\vec{\omega} = 0. \qquad (22)$$

Equation (22) describes shear wave propagation. The propagation velocity of the shear waves is

$$V_s = \sqrt{\frac{N}{\rho_{11}\left(1 - \frac{\rho_{12}^2}{\rho_{11}\rho_{22}}\right)}}. \qquad (23)$$

Equation (23) may be transformed as follows:

$$\rho_a^2 = \frac{\rho_{22}(\rho_{11}V_s^2 - N)}{(n+1)^2 V_s^2}. \qquad (24)$$

The function above relates $V_s$ to $\rho_a$. The coefficient $\rho_a$ is a measure of the inertial coupling between the fluid and the solid in the aggregate. Using equations (15) and (16), we obtain

$$\rho_a = \frac{\left(\rho_1+\rho_2-\frac{N}{V_S^2}\right)+\sqrt{\left(\frac{N}{V_S^2}-\rho_1-\rho_2\right)^2-4(n^2+2n)\left(\frac{N}{V_S^2}-\rho_1\right)\rho_2}}{2(n^2+2n)}. \qquad (25)$$

Considering the physics of $\rho_a$ and L'Hopital's rule, we can reject one of the $\rho_a$ solutions to obtain a unique solution for $\rho_a$. Note that the radicand in Eq. (25) should be non-negative.

We explore the relations between $\rho_a$ and relevant parameters by using the following example: a type of porous biomaterial for potential applications, poly(glycerol sebacate) (PGS), is the solid, and tissue fluid is the saturated fluid. We study two representative stages using two groups of data: the initial stage (group I), which commences upon implantation of porous PGS, and the final stage (group II), wherein tissue in-growth is near completion and PGS has been largely degraded. Table 1 lists the input parameters to our model.

**Table 1. Input parameters to model**

| Parameter | $\rho_s$ | $\rho_f$ | $N$ | $\emptyset$ |
|---|---|---|---|---|
| Group I | 1130 kg/m³ (Pomerantseva et al., 2009) | 1060 kg/m³ (Duck, 1990) | 0.05 MPa (Rai et al., 2012) | 0.3 |
| Group II | 1100 kg/m³ (Duck, 1990) | 1060 kg/m³ (Duck, 1990) | 0.32 MPa (Duck, 1990) | 0.03 |

In Table 1, $\rho_s$ corresponds to the mass density of the PGS bulk material in group I and the mass density of the cartilage (Xuan et al., 2020) in group II. The mass density of the tissue fluid in both groups is denoted by $\rho_f$. The term $N$ is used to denote the shear modulus of the porous PGS material in group I and the shear modulus of the cartilage in group II. The term $\emptyset$ denotes the aggregate porosity, which is used to calculate $\rho_1$ and $\rho_2$ from $\rho_s$ and $\rho_f$. The parameter $n$ is used to describe the relative motion between the fluid and the solid. From physical considerations, we consider that $\rho_f$ ranges from 0.1 to 0.5 in our analysis. The shear wave velocity, $V_s$, is considered to range from 1 m/s to 6 m/s.

We first study the condition that must be satisfied for Eq. (25) to have a physical meaning. First, we analyze the radicand in Eq. (25) to determine appropriate ranges for the relevant parameters. The radicand must be non-negative for $\rho_a$ to be real-valued. The two terms, $\left(\frac{N}{V_s^2}-\rho_1-\rho_2\right)^2$ and $4(n^2+2n)\left(\frac{N}{V_s^2}-\rho_1\right)\rho_2$, are plotted and compared to determine whether the inequality $\left(\frac{N}{V_s^2}-\rho_1-\rho_2\right)^2 - 4(n^2+2n)\left(\frac{N}{V_s^2}-\rho_1\right)\rho_2 \geq 0$ holds for group I and group II, respectively. Then, we graph the relation between $\rho_a$ and the relevant parameters to determine the dependence and sensitivity of certain variables, which serves as a basis for potential uses of the current model for evaluating the status of tissue in-growth.

## 3. RESULTS AND DISCUSSION

Figure 1 is a comparison of $\left(\frac{N}{V_s^2}-\rho_1-\rho_2\right)^2$ and $4(n^2+2n)\left(\frac{N}{V_s^2}-\rho_1\right)\rho_2$ for group I. When the shear wave velocity does not exceed 4 m/s, $\left(\frac{N}{V_s^2}-\rho_1-\rho_2\right)^2 \geq 4(n^2+2n)\left(\frac{N}{V_s^2}-\rho_1\right)\rho_2$ is guaranteed for all $n$= 0.1 ~ 0.5. Thus, the shear wave velocity can range from 1 m/s to 4 m/s, which is in accordance with typical applications (Taljanovic et al., 2017). A smaller $n$ tends to allow a wider range of shear wave velocities, which is mainly reflected in a higher upper limit for $V_s$.

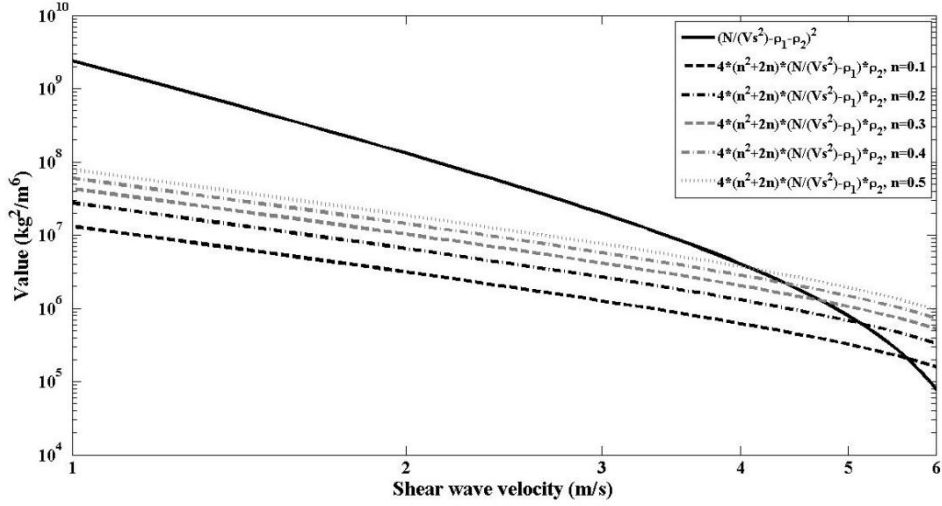

Fig. 1. Comparison of $\left(\frac{N}{V_s^2}-\rho_1-\rho_2\right)^2$ and $4(n^2+2n)\left(\frac{N}{V_s^2}-\rho_1\right)\rho_2$ for group I

In Fig. 2, the dependence of the inertial coupling parameter $\rho_a$ on $n$ and $V_s$ is plotted for group I, considering 1 m/s to 4 m/s as the appropriate range for $V_s$. Four plots corresponding to different shear wave speeds are shown. For a given $V_s$, as $n$ decreases, the value of $\rho_a$ increases. This variation is less evident at small $V_s$. The largest variation in $\rho_a$ in Fig. 2 is obtained at $V_s = 4$ m/s, which is presented as a dotted line. Thus, the theoretical model is most applicable to tissue repair processes with a shear wave velocity of 4 m/s, at which the $\rho_a$ response is most sensitive. The initial stage is simulated using group I, wherein the calculated $\rho_a$ is in the negative few hundred $kg/m^3$ range for an initial porosity of 0.3.

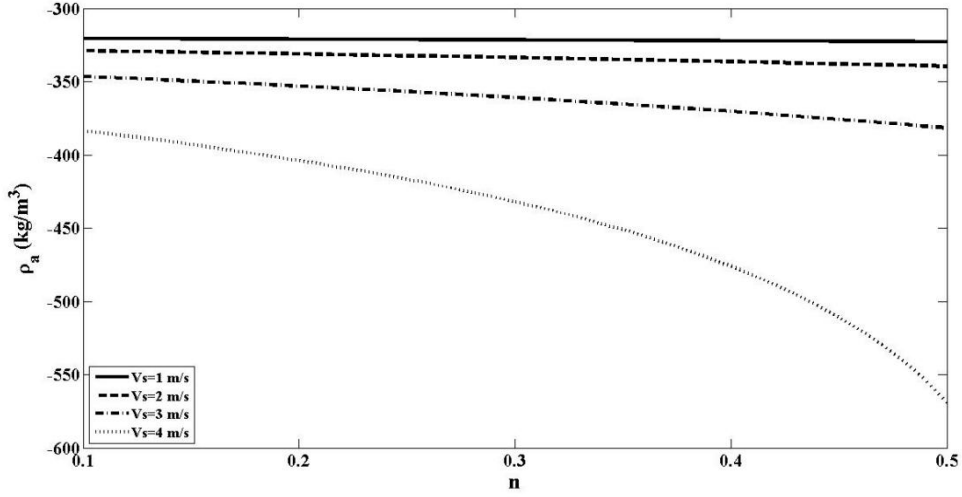

Fig. 2. Dependence of inertial coupling parameter $\rho_a$ on $n$ and $V_s$ for group I

The final stage, modeled by group II, is similarly graphed. Comparing with group I, the most significant changes are values of input parameters $N$ and $\emptyset$. To obtain a real-valued $\rho_a$, we once more consider the radicand in Eq. (25) and plot the corresponding two terms in Fig. 3. In the figure, $\left(\frac{N}{V_s^2}-\rho_1-\rho_2\right)^2$ is clearly greater than $4(n^2+2n)\left(\frac{N}{V_s^2}-\rho_1\right)\rho_2$ over the entire range of shear wave speeds from 1 m/s to 6 m/s, for all $n = 0.1 \sim 0.5$. A wider range of allowed $V_s$ is obtained for the final stage than in the initial stage using group I. However, for comparison with the group I results, we plot Eq. (25) in Fig. 4 using the previous $V_s$ range, namely, from 1 m/s to 4 m/s.

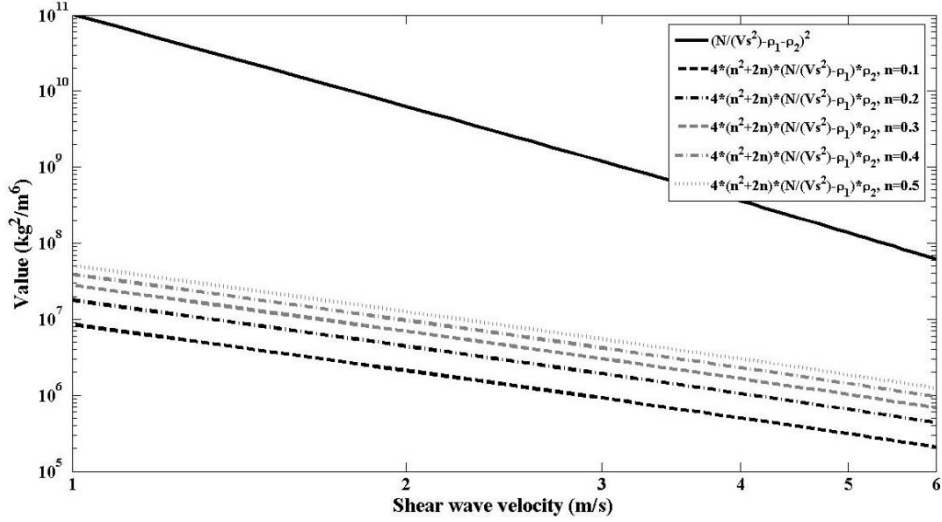

Fig. 3. Comparison of $\left(\frac{N}{V_s^2} - \rho_1 - \rho_2\right)^2$ and $4(n^2 + 2n)\left(\frac{N}{V_s^2} - \rho_1\right)\rho_2$ for group II

Figure 4 displays the dependence of $\rho_a$ on $n$ and $V_s$ for group II. A similar trend is observed as in Fig. 2, that is, for a given $V_s$, as $n$ decreases, $\rho_a$ increases. The most sensitive $\rho_a$ response is obtained at $V_s = 4$ m/s, but the difference among the four plots is not as distinct as in Fig. 2. The overall ranges of $\rho_a$ in the four plots are narrower than those in Fig. 2. In addition, the values of $\rho_a$ in each plot in Fig. 4 are closer to zero than in Fig. 2. This result is as expected. As tissue repair proceeds, $n$ approaches zero. From Eq. (24), we can derive $\rho_a = -\frac{\rho_2}{1 - \frac{\rho_2}{\frac{N}{V_s^2} - \rho_1}}$. As tissue in-growth proceeds, the porosity $\emptyset$ approaches zero: hence, $\rho_2$ approaches zero from Eq. (6), and $\rho_a$ approaches zero eventually. The model analysis and validation using the group I and group II data lead us to conclude that $\rho_a$ becomes less negative as tissue develops. A value of $\rho_a$ closer to zero corresponds to higher tissue recovery: therefore, $\rho_a$ could be utilized to evaluate the tissue remodeling process.

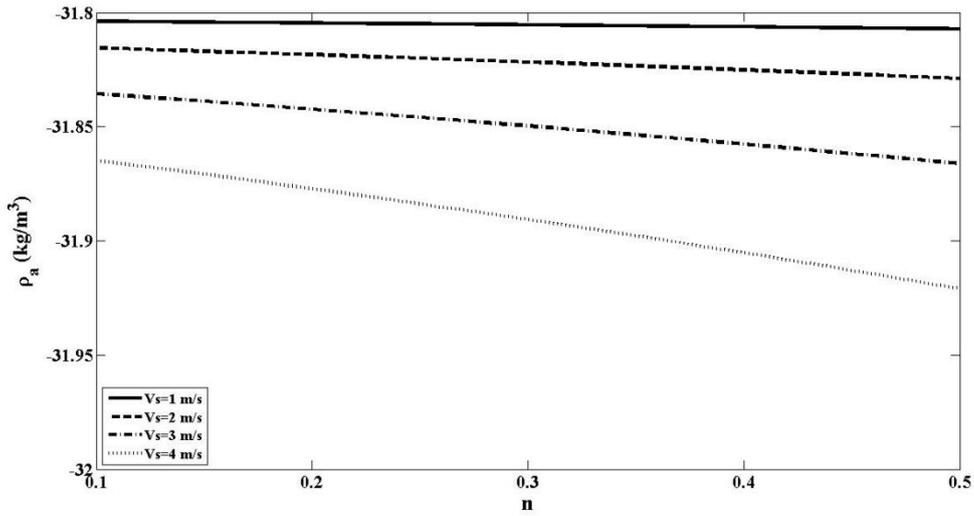

Fig. 4. Dependence of inertial coupling parameter $\rho_a$ on $n$ and $V_s$ for group II

The mathematical model for fluid-saturated porous bioelastomers formulated in the framework above can be utilized to describe the tissue remodeling process in tissue engineering. In our earlier research (Yu et al., 2013), quasi-static ultrasound elasticity imaging was used to observe the mechanical remodeling process along with tissue regeneration and porous bioelastomers degradation. This technique can be used as a nondestructive tool for imaging mechanical properties, but there are some limitations that affect its use, e.g. the technique relies on operator skill, and imaging accuracy can be affected by boundary conditions. In contrast, ultrasound shear wave elastography uses focalized ultrasound pulses to locally displace tissue and induce shear waves instead of using transducer pressure, and enables the measurement of the propagation speed of shear waves in the tissue to quantify the stiffness locally which is highly reproducible. Therefore, our objective is to provide theoretical guidance that can be used to quantitatively model and evaluate the tissue remodeling process.

Dating back to 1956, Biot (Biot, 1956a; Biot, 1956b) proposed a phenomenological theory for the propagation of elastic waves in a fluid-saturated porous solid, which is macroscopically isotropic and homogeneous. The theory was proved by Plona's experimental measurements in 1980 (Plona, 1980). In Biot's theory, considerable effort was expended to determine the two unknowns, that is, the speeds of shear waves and compressional waves. Finally, Biot obtained formulas for these two speeds. Our current research focus is the application of ultrasound shear wave elastography (Sarvazyan et al., 1998; Gennisson et al., 2013), which was developed in 1998, to evaluate the tissue remodeling process in the context of tissue engineering. Ultrasound shear wave elastography can be used to directly measure the shear wave speed. The shear wave speed can then be used to solve for the value of $\rho_a$, which is a critical indicator of the development of the tissue remodeling process. Thus, $\rho_a$ can be obtained using our developed method, in contrast to earlier studies (Biot, 1962; Plona et al., 1991), wherein $\rho_a$ was considered too complex to determine.

Our research has potential application in the field of tissue engineering, that is, ultrasound shear wave elastography could be used to investigate and evaluate the tissue remodeling process. Ultrasound shear wave elastography is used to obtain the shear wave speed, from which $\rho_a$ can be calculated to evaluate tissue recovery. Although the model in this paper is developed for soft bioelastomers, dissipation has not been considered in the present study. In our research on fluid-saturated porous bioelastomers, our current focus is the high-porosity initial stage and the low-porosity final stage. However, the dynamic parameters, e.g. $\emptyset$, $n$ and their corresponding relations, cannot be neglected in the biomechanical context. We will investigate these parameters more thoroughly using experimental measurements and statistical analysis in a later paper. Multiphase porous media is a highly complex subject with important technical applications, including to biomechanics. Biomechanical models are required to provide guidance for the non-invasive monitoring of the tissue remodeling process, whereby elucidation of the coupling process can be used to extract important observation parameters. To this end, we plan to apply the developed model in a pilot process evaluation in future studies.

## 4. CONCLUSION

In conclusion, we assume relative motion between a fluid and a fluid-saturated porous solid to develop a model to evaluate the fluid-solid coupling state in the tissue remodeling process. In the model, the shear wave velocity is used to obtain the coupling coefficient $\rho_a$, which is a measure of tissue recovery. This coefficient becomes increasingly less negative as tissue develops and approaches zero at the ideal final stage of zero porosity. Ultrasound shear wave elastography in combination with the developed model is proposed as a potential detection method for the regenerative process of engineered tissues. The model is tested for the initial and final stages of an implanted bioelastomer, and a shear wave velocity of 4 m/s is found to provide the most sensitive response of $\rho_a$. The developed method is anticipated to provide further insights into the field of acellular in situ tissue engineering for the non-invasive monitoring of the mechanical remodeling process of organ and tissue repair.

ACKNOWLEDGEMENTS

This work was financially supported by the National Natural Science Foundation of China (21574019, 11304137), the Natural Science Foundation of Liaoning Province of China (2019-MS-219), Liaoning Revitalization Talents Program (XLYC1907034), the Natural Science Foundation of Shanghai (18ZR1401900), and the Fundamental Research Funds for the Central Universities, DHU Distinguished Young Professor Program (LZA2019001).